\documentclass[aps,prl,preprint,groupedaddress]{revtex4-1}
\usepackage{multirow}
\usepackage{graphicx}
\usepackage{amsmath}
\usepackage{slashed}
\newcommand{\gosam}{\texttt{GoSam}}
\newcommand{\helacdipoles}{\texttt{HELAC-Dipoles}}
\newcommand{\helaconeloop}{\texttt{HELAC-Oneloop}}
\newcommand{\helacnlo}{\texttt{HELAC-NLO}}
\newcommand{\helacphegas}{\texttt{HELAC-PHEGAS}}

\newcommand{\lhapdf}{\texttt{LHAPDF}}
\newcommand{\madevent}{\texttt{MADEVENT}}

\newcommand{\powhel}{\texttt{PowHel}}
\newcommand{\powhegbox}{\texttt{POWHEG-Box}}

\newcommand{\pb}{\ensuremath{\,\mathrm{pb}}}

\newcommand{\mev}{\ensuremath{\,\mathrm{MeV}}}
\newcommand{\gev}{\ensuremath{\,\mathrm{GeV}}}
\newcommand{\tev}{\ensuremath{\,\mathrm{TeV}}}

\newcommand{\ep}{\epsilon}

\newcommand{\ttjet}{t\,$\bar{{\rm t}}$ + jet }
\newcommand{\pT}{\ensuremath{p_{\perp}}}

\newcommand{\pTZ}{\ensuremath{p_{\perp,Z}}}

\newcommand{\bt}{\ensuremath{\bar{{\rm t}}}}

\newcommand{\ttA}{t$\bar{{\rm t}}\gamma$}
\newcommand{\ttZ}{t$\bar{{\rm t}}Z$}
\newcommand{\ttH}{t$\bar{{\rm t}}H$}

\newcommand\Ref[1]     {Ref.\,\cite{#1}}

\newcommand\fig[1]     {Fig.\,{\ref{#1}}}

\begin{document}


\title{Top quark pair production in association with a $Z$-boson
at NLO accuracy}

\author{A.~Kardos}
\affiliation{Institute of Physics, University of Debrecen,
H-4010 Debrecen P.O.Box 105, Hungary}
\affiliation{Institute of Nuclear Research of the Hungarian
Academy of Sciences, Hungary}
\email[A.~Kardos: ]{adamkardos@inp.demokritos.gr}
\author{C.G.~Papadopoulos}
\affiliation{NCSR Demokritos, Institute of Nuclear Physics, Athens, Greece}
\author{Z.~Tr\'ocs\'anyi}
\affiliation{Institute of Physics, University of Debrecen,
H-4010 Debrecen P.O.Box 105, Hungary}
\affiliation{Institute of Nuclear Research of the Hungarian
Academy of Sciences, Hungary}

\date{\today}

\begin{abstract}
We present predictions for the production cross section of t-quark pair
production in association with a $Z$ boson at the next-to-leading
order (NLO) accuracy using matrix elements obtained from the
\helaconeloop\ package. We use the subtraction method for computing the
radiative corrections as implemented in the \powhegbox, which was also
used in several other computations of similar complexity. 
\end{abstract}

\pacs{12.38.-t, 13.87.-a, 14.65.Ha, 14.70.Hp}

\maketitle


In the recent years we have been witnessing the {\em NLO revolution} in
computing QCD jet cross sections. Before the millenium, one-loop
amplitudes were computed analytically, which was a serious bottleneck for
computing QCD radiative corrections to processes of final states with
high multiplicity. Indeed, the most complex computations involved
three-jet production with hadrons in the initial state
\cite{Nagy:2001xb,Nagy:2003tz} and electron-positron annihilation into
four-jets \cite{Nagy:1998bb}. The emergence of unitarity-based, fully
numerical approaches to computing one-loop amplitudes 
\cite{Bern:1994cg,Brandhuber:2005jw,Anastasiou:2006gt,Ossola:2006us,
Ellis:2007br,Bern:2007dw,Ossola:2008xq,Draggiotis:2009yb} has
changed the scope of possible computations completely. By now there are
publicly available computer programs for this purpose
\cite{Ossola:2007ax,Mastrolia:2010nb}, which paved the way to the
automation of NLO computations \cite{Hirschi:2011pa}. 

In this letter we present predictions for the process $pp \to$ \ttZ\ at
the NLO accuracy in QCD. This process is interesting from the point of view of
measuring the \ttZ\ couplings at the LHC \cite{Baur:2005wi}, which has
not yet been measured directly. This coupling is predicted by the
Standard Model (SM). Significant deviation from this predicted value
could be possible signal of new physics beyond the SM. In order to
optimize such measurements precise theoretical predictions are needed
both for the signal and the background processes.

Our present computations constitute steps in an ongoing project for
generating event samples for $pp \to$ t \bt $+X$ processes, where $X$ is
a hard partonic object in the final state
\cite{Kardos:2011qa,Garzelli:2011vp}. These event files are stored
according to the Les Houches accord \cite{Alwall:2006yp}, and can be
interfaced to standard shower Monte Carlo programs to produce
predictions for distributions at the hadron level that are exact up to
NLO accuracy upon expansion in the strong coupling. With such
predictions at hand one can optimize the selection cuts for the signal
process for improved experimental accuracy of the coupling measurements.

In our project we use the \powhegbox\ \cite{Alioli:2010xd}, which
produces a general framework also to NLO computations based
on the FKS subtraction scheme \cite{Frixione:1995ms}. The box requires
the relevant matrix elements as external input. For this purpose we use
the \helacnlo\ package \cite{Bevilacqua:2010mx,Bevilacqua:2011xh}, from
which we obtain the matrix elements in a semi-automatic way. Although
these tools have already been used successfully for many computations,
the complexity of the problem requires tedious checking and validation
procedures to make sure that the event samples can be safely used in
experimental analyses. Part of this procedure is a careful check
of the predictions at the NLO accuracy which is the subject of this
letter.


We performed our calculations using the \powhegbox\,
which requires the following ingredients:
\begin{itemize}
\itemsep -2pt
\item
Flavor structures of the Born ($gg\rightarrow Z t\bar{t}$,
$q\bar{q}\rightarrow Z t\bar{t}$, $\bar{q}q\rightarrow Z t\bar{t}$)
and real radiation ($q\bar{q}\rightarrow Z t\bar{t}g$,
$g g\rightarrow Z t\bar{t}g$, $\bar{q} g\rightarrow Z t\bar{t}\bar{q}$,
$g \bar{q}\rightarrow Z t\bar{t}\bar{q}$, $\bar{q}q\rightarrow Z t\bar{t}g$,
$q g\rightarrow Z t\bar{t}q$, $g q\rightarrow Z t\bar{t}q$)
subprocesses ($q\in\{u,d,c,s,b\}$).
\item
The Born-level phase space was obtained by using the invariant mass of
the t\bt-pair and four angles.
\item
Crossing invariant squared matrix elements with all incoming momenta for
the Born and the real-emission processes were built using amplitudes
obtained from \helaconeloop\ \cite{vanHameren:2009dr}
and \helacphegas\ \cite{Cafarella:2007pc}, respectively.  The matrix
elements in the physical channels were obtained by crossing.
\item
The color matrices for the color-correlated squared matrix elements
were taken from \helacdipoles\ \cite{Czakon:2009ss}.
\item
We used the polarization vectors to project the helicity amplitudes to
Lorentz basis for writing the spin-correlated squared matrix elements.
\end{itemize}
With this input \powhegbox\ can be used to perform the necessary
integrations numerically. In order to ensure the correctness of the
computations, we performed comparisons of the matrix elements to those
obtained from independent sources and consistency checks.

We compared the tree-level (Born and real-emission) matrix elements to
those obtained from \madevent\ \cite{Maltoni:2002qb}, while the matix
element for the virtual correction, including the first three terms in
the Laurent expansion in the dimensional regularisation parameter $\ep$
was compared to that from \gosam\ \cite{Mastrolia:2010nb} in randomly
chosen phase space points.  In all cases we found agreement up to at
least 6 digits.  We also computed various distributions at leading
order with both \powhel\ (=\powhegbox+\helacnlo) and \madevent\ and
found agreement.  

The consistency between real-emission, Born, color-correlated and
spin-correlated matrix elements was checked by taking the soft- and
collinear limits of the real-emission squared matrix elements in all
possible kinematically degenerate channels using randomly chosen phase
space regions. As a further consistency check, we have also implemented
the similar process $pp \to$ \ttA, which has also been computed at the
NLO accuracy \cite{Duan:2009zza,Melnikov:2011ta}, and we found
agreement with the predictions of \Ref{Melnikov:2011ta}. The important
difference from the point of view of the computations lies in the mass
of the vector boson. In this respect the \ttA-production resembles
\ttjet-production presented in \Ref{Kardos:2011qa}, while \ttZ-production
resembles \ttH-production, presented in \Ref{Garzelli:2011vp}.


For \ttZ\ hadroproduction the fully inclusive cross section and the
distribution of the transverse momentum of the $Z$-boson is the only
published prediction at NLO accuracy. In \Ref{Lazopoulos:2008de} the
prediction was made with the default scale choice $\mu (= \mu_R = \mu_F) =
\mu_0 (= m_t + m_Z/2)$.  With this scale and using the same input
parameters we find agreement for the LO prediction (0.808\pb), including
the dependence on the scale $\mu$. However, at NLO accuracy our
prediction is slightly larger, (1.121$\pm$0.002)\pb\ compared to
1.09\pb\ \cite{Petriello:private} in \Ref{Lazopoulos:2008de}, leading
to an inclusive $K_{\rm inc}=1.39$ (instead of the $K_{\rm inc} = 1.35$
in \Ref{Lazopoulos:2008de}).

\begin{figure}
\includegraphics[width=8cm]{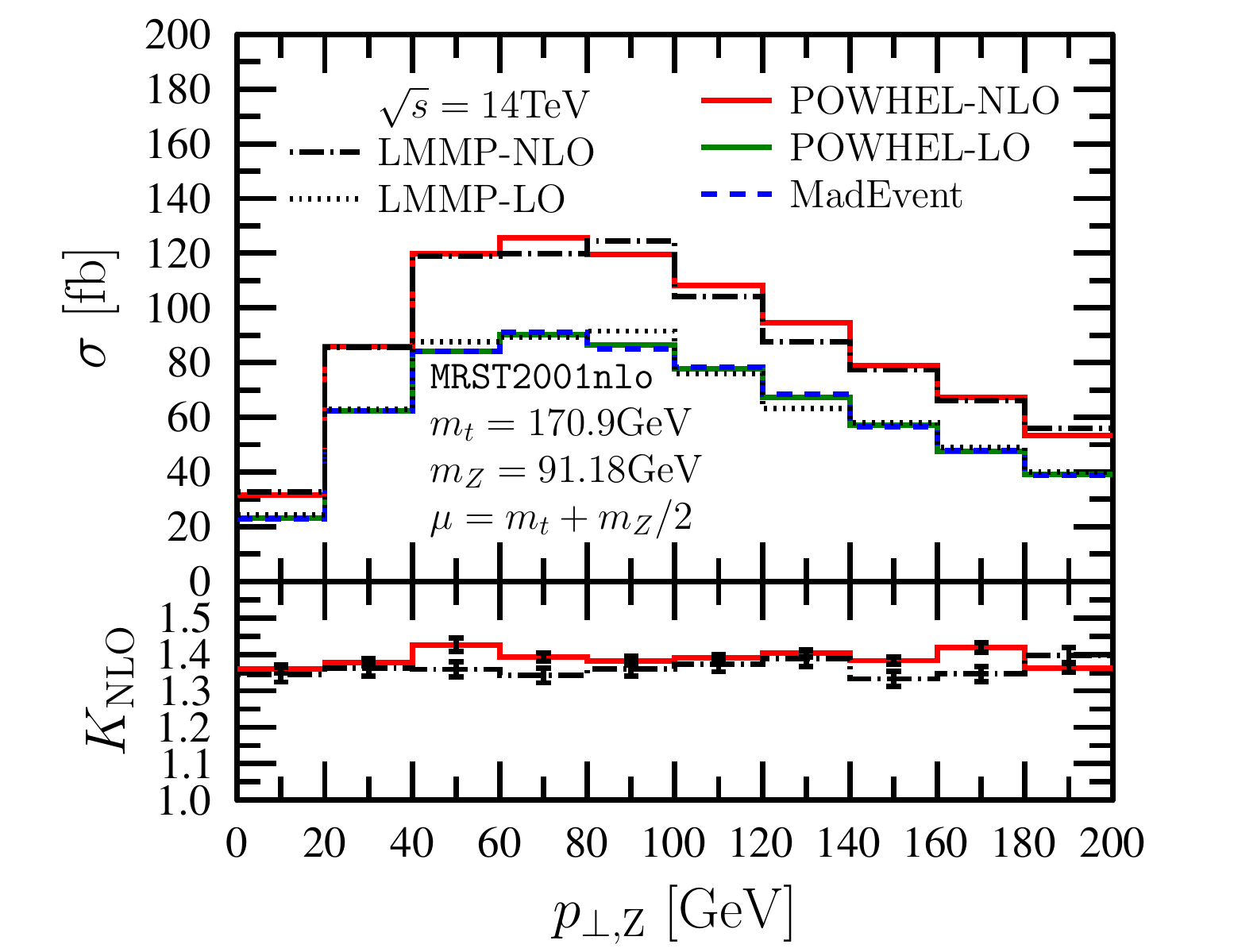}
\caption{Transverse momentum distribution of the $Z$-boson 
from \Ref{Lazopoulos:2008de} (LMMP) and from our calculation
(PowHel). The lower panel shows our prediction for the NLO
$K$-factor compared to that of LMMP.}
\label{fig:compare_pTZ}
\end{figure}

For the \pTZ-distribution we checked our LO prediction with \madevent,
and found agreement as shown in \fig{fig:compare_pTZ}, but not with
the LO predictions of \Ref{Lazopoulos:2008de}. As a result, the
predictions of the two computations disagree also at NLO. In order to
separate the effect of the differences at LO, in the lower panel we
plot the $K$-factor of our computation against the 
$K_{\rm NLO}$ of \Ref{Lazopoulos:2008de}. The two predictions agree
within the uncertainties of the numerical integrations except in two
bins.

\begin{figure}
\includegraphics[width=8cm]{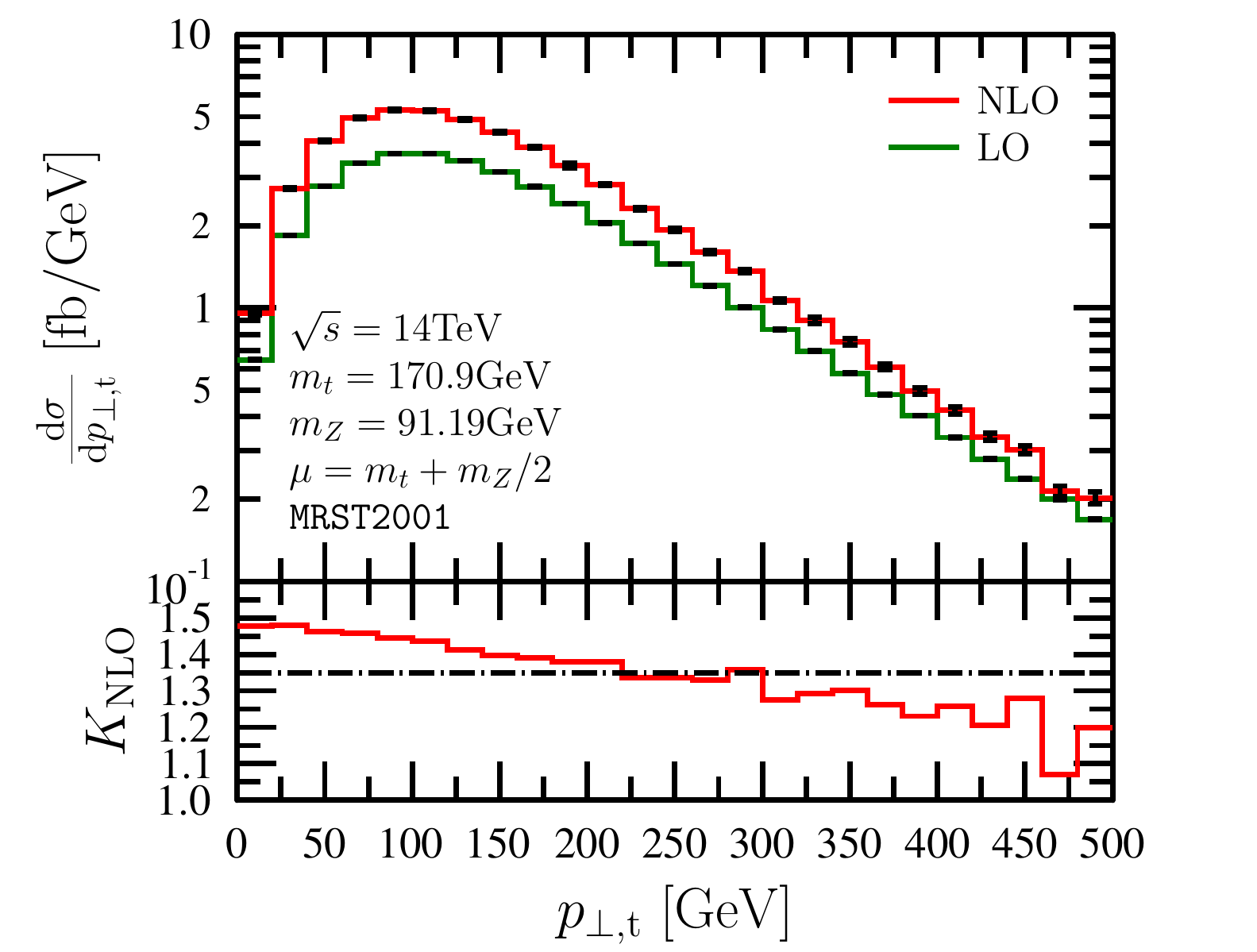}
\caption{Transverse momentum distribution of the t-quark. The lower
panel shows our prediction for the NLO $K$-factor compared to a
constant value $K_{\rm inc} = 1.39$.}
\label{fig:pTt}
\end{figure}

The almost constant value of the $K$-factor for the \pT-distribution of
the $Z$-boson made the authors of \Ref{Lazopoulos:2008de} expect that the
shape of many other kinematic distributions will also be approximately
unchanged by NLO corrections. Our findings do not support this
expectation. As an example, we show the \pT-distribution of the t-quark
in \fig{fig:pTt}.

Next we turn our attention to make predictions for \ttZ\
hadroproduction at the current LHC.
The parameters of our calculation were the following: $\sqrt{s} = 7\tev$,
\texttt{CTEQ6.6M} PDF set \cite{Nadolsky:2008zw} from \lhapdf{}, 2-loop
running $\alpha_s$, with $\Lambda_5^{\overline{\mathrm{MS}}} =
226\mev$, $m_t = 172.9\gev$, $m_Z = 91.1876\gev$, $m_W = 80.399\gev$,
$G_F = 1.16639\cdot 10^{-5}\gev^{-2}$, the renormalization and
factorization scales were chosen equal to $\mu_0 = m_t + m_Z/2$. 

\begin{figure*}
$
\begin{array}{cc}
  \includegraphics[width=8cm]{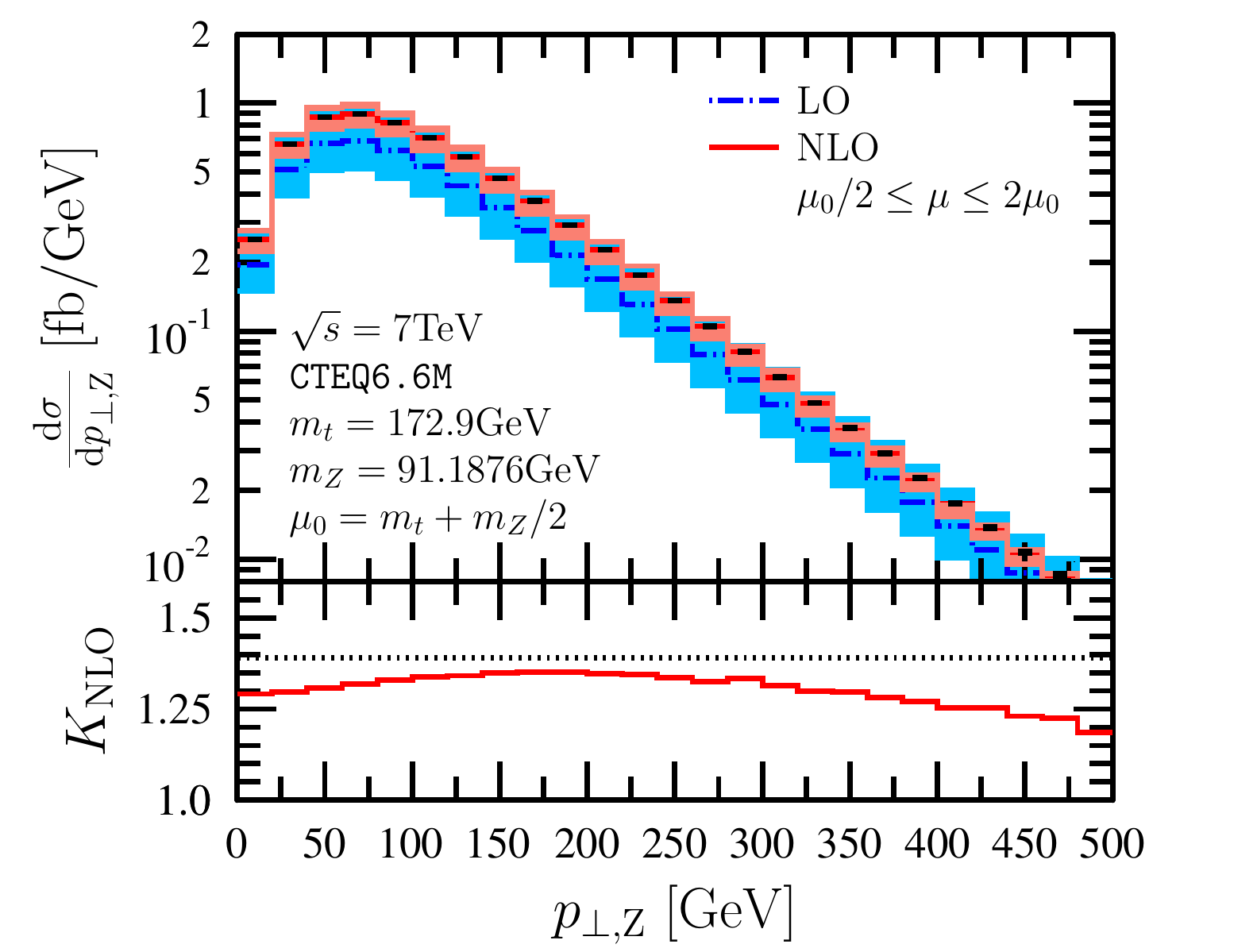}
& \includegraphics[width=8cm]{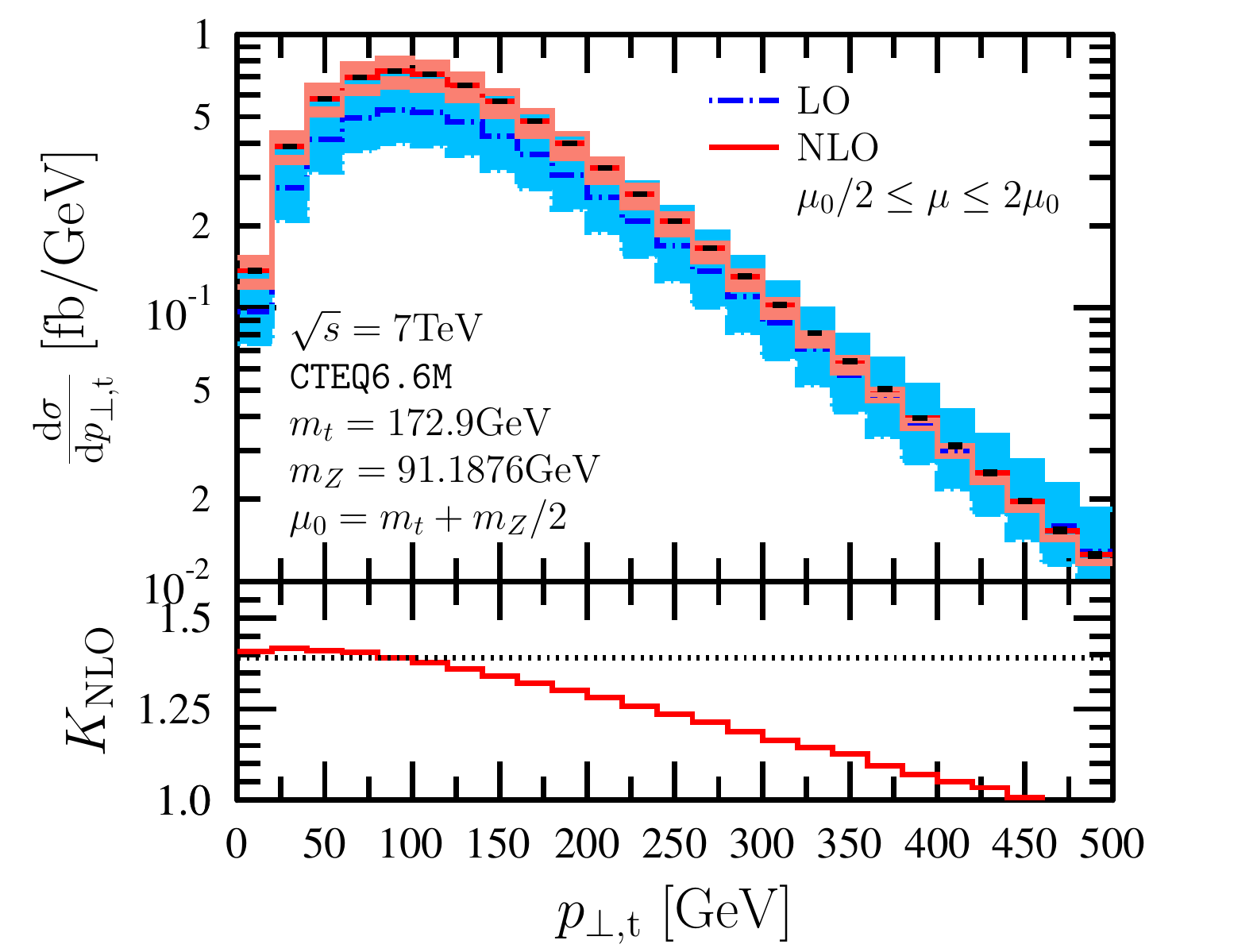}
\end{array}
$
\caption{Transverse momentum distributions of the $Z$-boson (left) and
the t-quark (right). The lower panels show our prediction for the NLO
$K$-factors compared to the constant value $K_{\rm inc} = 1.39$ found at
14\tev\ and default scale $\mu_0$.}
\label{fig:pt_LHC}
\end{figure*}

In \fig{fig:pt_LHC} we show the transverse momentum distributions of the
$Z$-boson and the t-quark. We find that the $K$-factor for the
\pTZ-distribution is somewhat smaller and less uniform than that at
14\,TeV. The bands correspond to the simultaneous variation of the
renormalization and factorization scales between one half and two times
the default value. We observe a significant reduction of the scale
dependence as going from LO to NLO, also seen for other distributions.
The extra radiation, present in the NLO computation, makes the spectra
softer as expected.  This softening decreases the $K$-factor below one
for very large values of transverse momentum of the t-quark. Around the
value of \pT\ where the width of the NLO band shrinks to zero
accidentally, and the band is unlikely to represent the effect of the
missing higher order contributions.  

\begin{figure*}
$
\begin{array}{cc}
  \includegraphics[width=8cm]{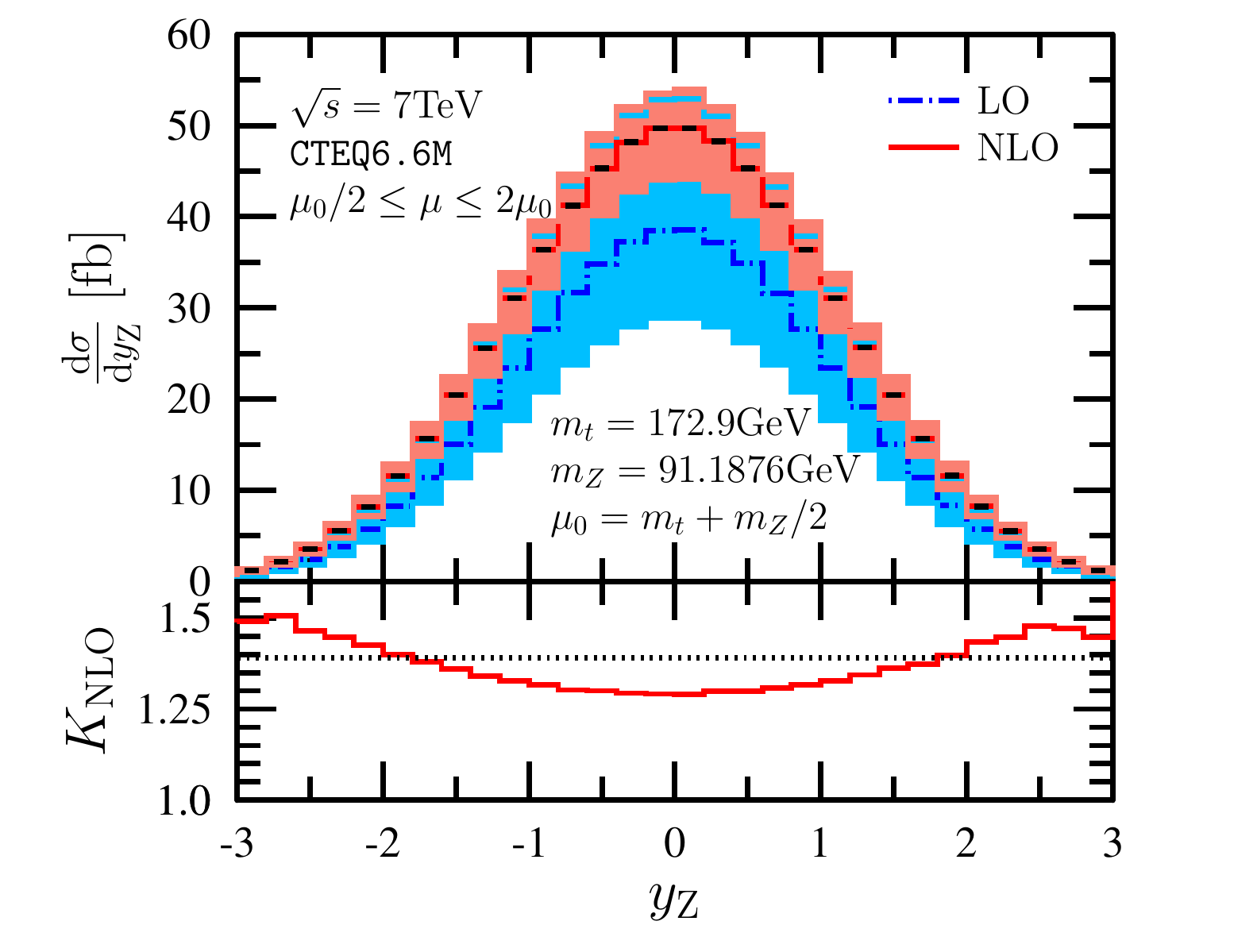}
& \includegraphics[width=8cm]{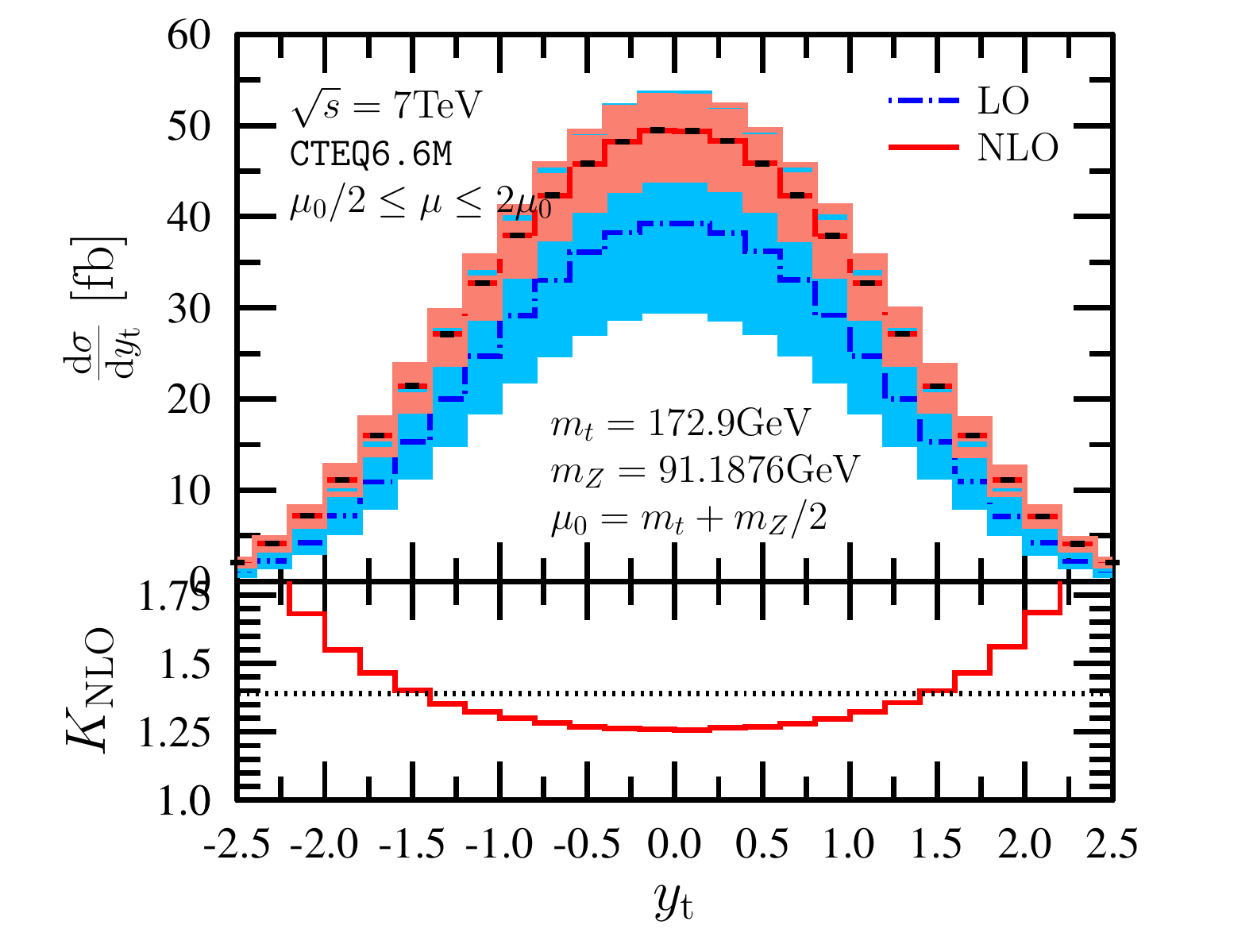}
\end{array}
$
\caption{Rapidity distributions of the $Z$-boson (left) and the t-quark
(right). The lower panels show our prediction for the NLO $K$-factors
compared to the constant value $K_{\rm inc} = 1.39$ found at 
14\tev\ and default scale $\mu_0$.}
\label{fig:y_LHC}
\end{figure*}

In \fig{fig:y_LHC} we show the rapidity distributions of the $Z$-boson
and the t-quark. The corrections are clearly not uniform, but increase
towards large rapidities, making the predictions less reliable in those
regions. For central rapidities the corrections are moderate and the
scale uncertainty decreases significantly.

\begin{figure*}
$
\begin{array}{cc}
  \includegraphics[width=8cm]{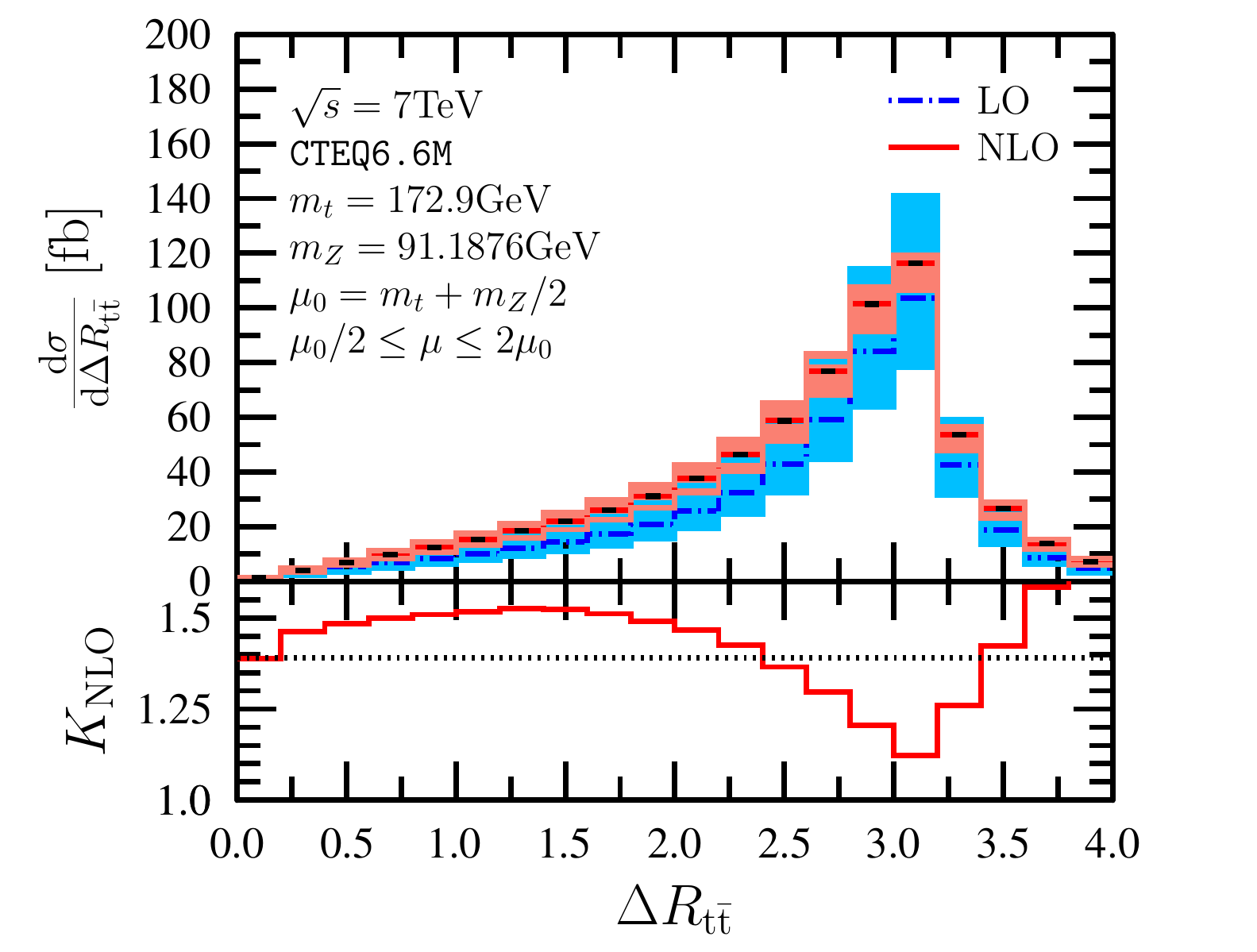}
& \includegraphics[width=8cm]{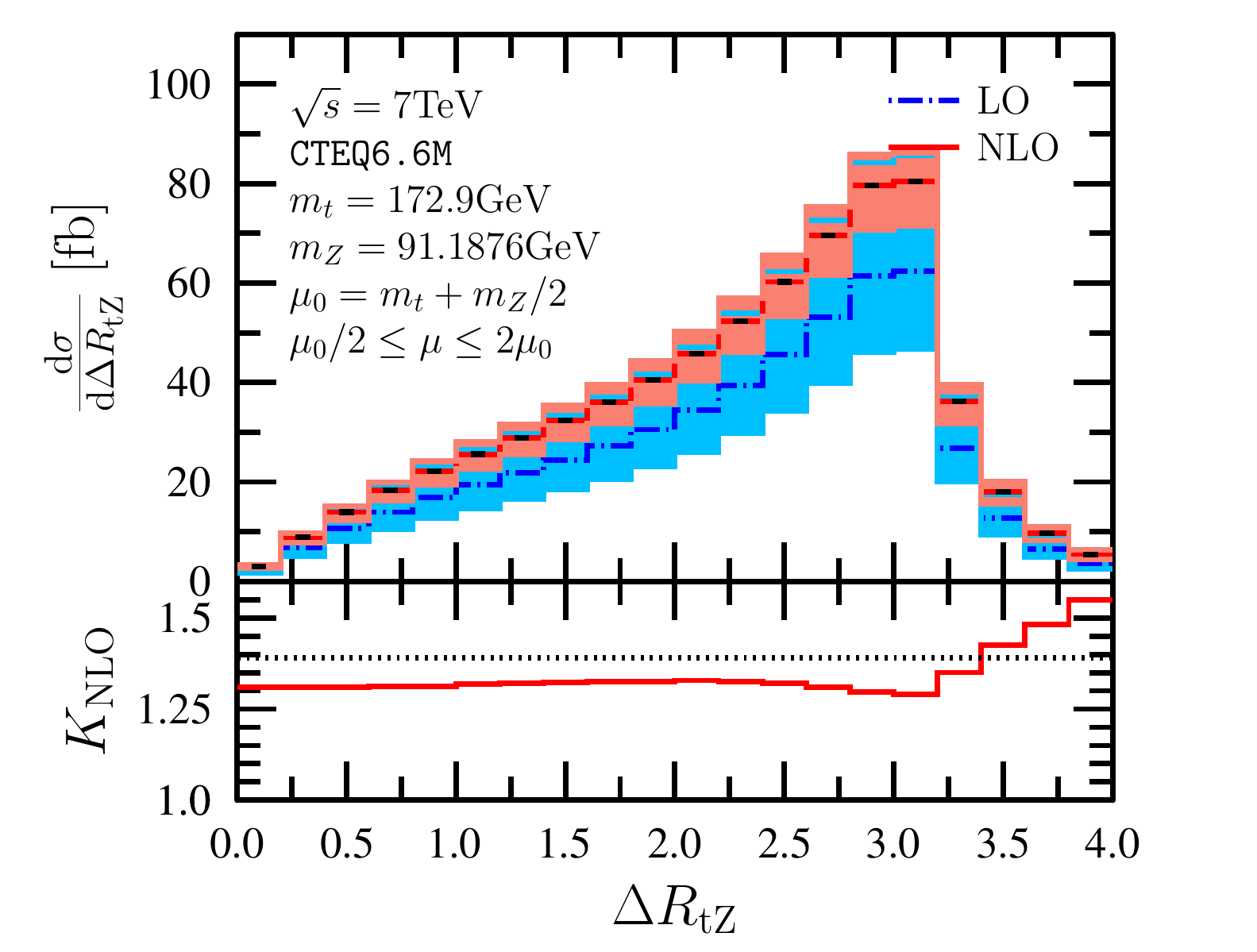}
\end{array}
$
\caption{Distributions of the $R$-separation between the t\bt-pair
(left) and between the t-quark and $Z$-boson (right). The lower panels
show our prediction for the NLO $K$-factors compared to the constant value
$K_{\rm inc} = 1.39$ found at 14\tev\ and default scale $\mu_0$.}
\label{fig:dr_LHC}
\end{figure*}

Finally, \fig{fig:dr_LHC} shows distributions of the separation in the
rapidity-azimuthal angle plane of the t\bt-pair and also of the t-quark
$Z$-boson. For the t\bt-pair the corrections are large except for the
back-to-back configuration, while for the t$Z$-separation the
corrections are moderate except for large rapidities.



In this letter we studied the hadroproduction of a t\bt-pair in
association with a $Z$ boson. This process is of interest for
measuring the \ttZ-coupling directly at the LHC. Our predictions
slightly disagree with the results of the only published computation.
The difference most likely originates from the lower numerical precision
of the integrations in \Ref{Lazopoulos:2008de}.

We produced predictions for the LHC. In general one can conclude that the
NLO predictions are sizeable, but not alarmingly large, lie in the
30--40\,\% range except towards the edges of the kinematically available
regions, where the $K$-factor can grow above 1.5. As a result the
dependence on the renormalization and factorization scales decreases
significantly and the theoretical prediction becomes fairly reliable.

In order that our predictions become useful for the measurement of the
\ttZ-couplings, matching with shower Monte Carlo programs is desirable,
which we consider in a separate publication. With such a matching we can
let the t-quarks decay and study the optimization of the experimental
cuts realistically.


This research was supported by
the HEPTOOLS EU program MRTN-CT-2006-035505,
the LHCPhenoNet network PITN-GA-2010-264564,
the Swiss National Science Foundation Joint Research Project SCOPES
IZ73Z0\_1/28079, 
and the T\'AMOP 4.2.1./B-09/1/KONV-2010-0007 project.
Z.T. thanks the Galileo Galilei Institute for Theoretical Physics for the
hospitality and the INFN partial support during the completion of this
work.
We are grateful
to F.~Tramontano for comparing our one-loop amplitudes to results of \gosam,
to F.~Petriello for clarifying the numerical uncertainties in
\Ref{Lazopoulos:2008de},
and to the NIIFI Supercomputing Center for access to their computers.

\end{document}